\documentclass[12pt]{article}
\usepackage{amsmath}
\usepackage{graphicx,psfrag,epsf}
\usepackage{enumerate}
\usepackage{natbib}
\usepackage{amsmath,amsthm,amsfonts,epsfig,amssymb,natbib,eucal,eufrak}
\usepackage{booktabs}
\usepackage{multirow}
\usepackage{siunitx}
\usepackage{qtree}

\usepackage{hyperref}

\usepackage{float}
\restylefloat{table}
\usepackage{color}

\usepackage[ruled,vlined,linesnumbered]{algorithm2e}

\newtheorem{remark}{Remark}

\newcommand{\blind}{0}

\begin{document}

\def\spacingset#1{\renewcommand{\baselinestretch}%
{#1}\small\normalsize} \spacingset{1}


\if0\blind
{
  \title{\bf An optimal design for hierarchical generalized group testing}
  \author{Yaakov Malinovsky
    \thanks{Corresponding author}\\
    Department of Mathematics and Statistics\\ University of Maryland, Baltimore County, Baltimore, MD 21250, USA\\
    \\
    Gregory Haber and Paul S. Albert \thanks{ The work was supported by the National Cancer Institute Intramural Program.}\hspace{.2cm}\\
    Biostatistics Branch, Division of Cancer Epidemiology and Genetics\\National Cancer Institute, Rockville, MD 20850, USA
    }

  \maketitle
} \fi

\if1\blind
{
  \bigskip
  \bigskip
  \bigskip
  \begin{center}
    {\LARGE\bf Title}
\end{center}
  \medskip
} \fi

\bigskip
\begin{abstract}
Choosing an optimal strategy for hierarchical group testing is an important problem for practitioners who are interested in disease
screening with limited resources. For example, when screening for infectious diseases in large populations, it is important to use algorithms that minimize the cost of potentially expensive assays.
\cite{BBT2015} described this as an intractable problem unless the number of individuals to screen is small.
They proposed an approximation to an optimal strategy
that is difficult to implement for large population sizes.
In this article, we develop an optimal design with respect to the expected total number of tests that can be obtained using a novel dynamic programming algorithm. We show that this algorithm is substantially more efficient than the approach proposed by \cite{BBT2015}. In addition, we compare the two designs for imperfect tests. R code is provided for the practitioner.
\end{abstract}

\noindent%
 \vfill

\newpage
\spacingset{1.45} 
\section{Introduction}
Screening populations for infectious diseases is important for early detection;
an example is screening for human papillomavirus (HPV) infection for the
early detection of
cervical cancer and its precursors \citep{N2000, SW2013}. In under-resourced countries, individual testing may be expensive and therefore not feasible.  The use of group testing where samples are combined in a single test can lead to cost savings. This paper proposes
an optimal design for this type of screening.

Group testing for identification when the probability of disease varies across subjects, which has been called the generalized group testing problem (GGTP), is a challenging problem in applied statistics
and was first introduced by \cite{S1960}.
Recently in this journal, \cite{BBT2015} introduced an algorithm for
the GGTP in a hierarchical class. A procedure is in the {\it hierarchical class} if two units are tested together in a group only if they have an identical test history, i.e., if each previous group test contains either both of them or none of them \citep{SG1959,HPE1981}. Additional non-optimal hierarchical algorithms for the GGTP have been developed \citep{LTP1994, BT2010, M2017}.

In this article, we develop an optimal hierarchical algorithm for GGTP that uses dynamic programming (DP) to find an optimal design with respect to the expected total number of tests.
R code is available that computes the optimal design and the associated expected total number of tests for a given population size
and individual-specific prevalences (https://github.com/habergw/genGT). We show that the proposed approach is substantially more efficient with respect to both the expected total number of tests and computational performance than the approach proposed by \cite{BBT2015} (CRC procedure).
Our focus is on an optimal hierarchical design for GGTP, where tests are error-free. A full discussion of the impact of using
the algorithm with imperfect tests is presented in the Appendix.

\section{An optimal Hierarchical Algorithm}
\label{se:OHDP}

We assume without loss of generality that  $p_1\leq p_2\leq \cdots \leq p_N$  with the corresponding labels $1,\ldots,N$,
where $N$ is the size of the population and $p_i$ is the known probability of an infection for each person in the population
(as in \cite{BBT2015}).
We develop an optimal hierarchical algorithm with respect to the ordered values of $p_i$.
This imposes the restriction that for any two subgroups, $p_i$ values  in one subgroup are all greater than or equal to every $p_i$ value in the other subgroup.
The ordering assumption is necessary in order to make the problem tractable since optimizing with respect to all possible permutations of $p_1,\ldots,p_N$ is impossible even for small $N$. Additional discussion regarding the issue of ordering can be found
in \cite{M2017} and references therein.

The DP algorithm begins by dividing the population $U=\left\{1,\ldots,N\right\}$ of $N$ units into $S$ subsets ($1\leq S\leq N$) $I_1,\ldots,I_S$.
The units in each subset are combined and tested together (stage 1).
The $S$ tests in stage 1 are binary tests where $X_{I_i}=1,\,\,i=1,\ldots,S$ if at least one subject in $I_i$ is positive, and $X_{I_i}=0$ otherwise. For $X_{I_i}=0$, we conclude that all subjects in $I_i$ are negative; otherwise, $I_i$ proceeds to stage 2.
In stage 2, we choose a proper subset $A_i$ from each $I_i$.
If $X_{A_i}=0$, we can infer that $X_{A^{c}_i}=1$ without testing $A^{c}_i$, where $A^{c}_i=I_i-A_i$ (since $X_{I_i}=1$).
Then we proceed to stage 3 where we will test an appropriate subset of $A^{c}_i$.
If $X_{A_i}=1$, we proceed to stage 3 where we test appropriate subset of $A_i$ and $A^{c}_i$ as a whole.
We proceed in a similar manner until the status of all units is known. Tests can be performed simultaneously at each stage.

The proposed hierarchical dynamic programming (HDP) algorithm differs from CRC in a number of ways. First, HDP explicitly computes the hierarchical design that minimizes the expected total number of tests. The CRC is not optimal in that sense.
Second, CRC requires the implicit specification of the first stage configuration, which is difficult to choose in a principled way.
The splitting of groups in subsequent stages for CRC is based on a heuristic search that is not necessarily optimal. In contrast, the HDP algorithm chooses the initial and subsequent groupings in an optimal way.  Third, the CRC allows for the specification of the maximum number of stages, up to 4.
In comparison, the HDP does not allow for the specification of the maximum number of stages apriori.
Last, as we will show in Section 4, CRC is substantially more computationally expensive than HDP.

For the HDP algorithm, the optimal configuration is found through dynamic programming that uses backward induction \citep{B1957,L1961}. We use a heuristic argument to explain the algorithm to practitioners (the Appendix provides technical details). First, we explicitly find an optimal design for two subjects with corresponding probabilities $p_{N-1}\leq p_{N}$. Then we find an optimal design for three subjects ($p_{N-2}\leq p_{N-1}\leq p_{N}$) using the optimal design previously found for two subjects. We proceed in this fashion through the entire population.

Although we delegate the technical details to an Appendix \ref{A:aa}, we need to introduce notation in order
for the practitioner to apply the algorithm. Let $\displaystyle {B_{n:N}=\left\{n,\ldots,N\right\}}$ (Binomial set) denote the set of units with labels $n,n+1,\ldots,N$ and corresponding probabilities $p_n,\ldots,p_N$.
By $H(n:N)$, we denote the expected number of tests under an optimal HDP algorithm applied to the set $B_{n:N}$.
The backward induction process starts with $H(N:N)=1$, and $H(n:N)$ is determined recursively for $n=N-1, N-2,\ldots, 1$.
In the process of testing, some groups will be identified as positive. Therefore, we introduce additional notation for the expected total number of tests under an optimal HDP algorithm conditional on the information that there is at least one infected individual in the group. Let $\displaystyle{D_{n:n_1}=\left\{n,\ldots,n_1\right\}}$ (Defective set) denote the set of such units with labels $n,\ldots,n_1$. We denote by $h(n:n_1)$ the expected total number of tests under an optimal HDP algorithm applied to the set $\displaystyle{D_{n:n_1}}$, where $\displaystyle{h(n:n)=0,\,\,\, n=1,\ldots,N}$. Let $\displaystyle{\Pi\left(a:b\right)=q_{a}q_{a+1}\cdots q_{b},\,\,\,q_i=1-p_i}$. The HDP algorithm is shown below.

\begin{algorithm}[H]
\DontPrintSemicolon
\SetAlgoLined
\SetKwInOut{Input}{Input}
\SetKwInOut{Initial}{Initial Values}
\SetKwInOut{Output}{Output}
\Input{$p_1\leq p_2\leq \cdots \leq p_N$}
\Initial{$H(N+1:N)=0,\,H(N:N)=1,\, h(n:n)=0\,\,for\,\, n=1,\ldots,N$}
\BlankLine

\For{n:=N-1 to 1 step 1  }{

    \For{k:=1 to N-(n-1) step 1}
    {\While {$k>1$}
       { \For{x=1: to k-1 step 1}{
        $
        T(n,k,x)=2-\frac{\Pi\left(n:n+(x-1)\right)\left(1-\Pi\left(n+x:n+(k-1)\right)\right)}{1-\Pi\left(n:n+(k-1)\right)}+
        \frac{1-\Pi\left(n:n+(x-1)\right)}{1-\Pi\left(n:n+(k-1)\right)}h(n:n+(x-1))+
        \frac{1-\Pi\left(n+x:n+(k-1)\right)}{1-\Pi\left(n:n+(k-1)\right)}h(n+x:n+(k-1))
        $
        \tcp*{the expected total number of test for the defective set $D_{n:n+(l-1)}$ if we first test $x$ units. See \ref{A2:h} for developments.}
        }
        $h(n:n+(k-1))=\min_{1\leq x \leq k-1}T(n,k,x)$
        \tcp*{optimal expected value. See \ref{A2:h} for developments. }}
        \;

        $k^{*}(n:n+(k-1))=\arg\min_{1\leq x \leq k-1}T(n,k,x)$
        \tcp*{From the Defective set $\left\{n,\ldots,n+(k-1)\right\}$ is optimal to test first $k^{*}$ items. }
        \bigskip

       $T_{k}(n:N)=1+\left\{H(n+k:N)+\left(1-\Pi\left(n:n+(k-1)\right)\right)h(n:n+(k-1))\right\}$
\tcp*{the expected total number of test for the binomial set $B_{n:N}$ if we first test $k$ units. See \ref{A1:H} for developments.}
    \;
    \;
    }
$H(n:N)=1+\min_{1\leq k \leq N-(n-1)}\left\{T_{k}(n:N)\right\}$
\tcp*{optimal expected value. See \ref{A1:H} for developments. }
\;
$k^{**}(n:N)=\arg\min_{1\leq k \leq N-(n-1)}\left\{T_{k}(n:N)\right\}$
\tcp*{From the Binomial set $\left\{n,\dots,N\right\}$ is optimal to test first $k^{**}$ items. }
    \;
}
\caption{An optimal HDP algorithm: design and the value of $H(1:N)$}
\label{A:1}
\end{algorithm}
// denote comment.

\noindent
In the next section, we illustrate the implementation of the HDP algorithm with a small population size.

\section{Demonstration of the HDP Algorithm}
We begin by demonstrating the HDP algorithm for two subjects $a$ and $b$,
 where
$q_a\geq q_b $ and $q_i=1-p_i$, $i=a,b$. Denote by $T$ the total number of tests.
The left branch of the tree represents the negative test result, and the right branch represents the positive test result.
\begin{center}
{\bf Algorithm $A_{a,\,b}$}
\scriptsize
\Tree [.{test together units a and b} [.{$T=1$ with prob. $q_aq_b$} ] [.{test unit a} [.{$T=2$ with prob. q_a(1-q_b)} ] [.{test unit b} [.{$T=3$ with prob. (1-q_a)q_b}  ][.{$T=3$ with prob. (1-q_a)(1-q_b)} ] ] ] ]
\end{center}

Using algorithm $A_{a,\,b}$, we show the implementation of the DP algorithm with probability vector $p=\left(0.05, 0.07, 0.09, 0.11, 0.13, 0.15, 0.17, 0.19, 0.21, 0.23, 0.25 \right)$ with corresponding labels set $\left\{1,2,\ldots,11\right\}$.
Through backward induction, the optimal configuration of the initial stage is $I_1=\left\{1,2,3,4,5,6\right\}, I_2=\left\{7,8,9\right\}, I_3=\left\{10,11\right\}$.
Subsequent testing is done separately in each of the three groups based on the HDP algorithm.
We present below three testing trees corresponding to the optimal initial configuration of three subgroups $I_1, I_2, I_3$.
Recall that the left branch of the tree represents the negative test result and the right branch represents the positive test result.
\bigskip

\noindent
{\bf subgroup} $I_1=\left\{1,2,3,4,5,6\right\}$:
\begin{center}
\qtreecenterfalse
{{
\Tree [.{test ${\left\{1, 2, 3, 4, 5, 6\right\}}$}
Stop [.{test $\left\{1,2,3\right\}$}
[.{test $\left\{4\right\}$ }
[.{test $\left\{5\right\}$} Stop  {test $\left\{6\right\}$} ] {test $\left\{5, 6\right\}$: algorithm $A_{5,\,6}$}
]
[.{\,\,\,\,\,\,test $\left\{1\right\}$ and follow A}
[.{test $\left\{2\right\}$}  Stop {test $\left\{3\right\}$} ]  {test $\left\{2, 3\right\}$: algorithm $A_{2,\,3}$}
] ] ]
]
}}
\end{center}
\bigskip

\begin{center}
\qtreecenterfalse
{{
\Tree [.{A: test ${\left\{4, 5, 6\right\}}$}
Stop [.{test $\left\{4\right\}$} [.{test $\left\{5\right\}$} Stop  {test $\left\{6\right\}$} ] {test $\left\{5, 6\right\}$: algorithm $A_{5,\,6}$}
] ]
}}
\end{center}

\noindent
{\bf subgroup} $I_2=\left\{7, 8, 9\right\}$:

\begin{center}
\qtreecenterfalse
{{
\Tree [.{test ${\left\{7, 8, 9\right\}}$}
Stop [.{test $\left\{7\right\}$} [.{test $\left\{8\right\}$} Stop  {test $\left\{9\right\}$} ] {test $\left\{8, 9\right\}$: algorithm $A_{8,\,9}$}
] ]
}}
\end{center}

\noindent
{\bf subgroup} $I_3=\left\{10, 11\right\}$:\\
test $\left\{10, 11\right\}$: algorithm $A_{10,\,11}$.
\medskip

\noindent
The expected number of tests under this optimal HDP design is 6.820.

\section{Numerical comparisons with \cite{BBT2015} and common used algorithms}
We compare the performance of the HDP algorithm with that of the CRC method proposed by \cite{BBT2015}.
CRC requires the specification of the maximum number of stages therefore, we used either three or four stages, depending on $N$.
We also compare the performance of HDP and CRC with those of two commonly used hierarchical algorithms:
a Dorfman-type \citep{Dorfman1943} algorithm and a sequential algorithm proposed by
\cite{S1957}. We abbreviate these algorithms as procedures $D^{'}$ and $S$, respectively. In the GGTP setting, the optimum ordered configuration for both algorithms was characterized by \cite{M2017}.
For the CRC, we obtained the expected total number of tests using R functions provided by \cite{BBT2015}.
The comparisons are done in the following manner.
We generate the vector $\displaystyle p_1,p_2,\ldots,p_{N}$ from a Beta distribution with parameters $\displaystyle \alpha=1, \beta>0$ such that
$\displaystyle \frac{1-p}{p}=\beta$, i.e., expectation equal to $p$ and repeat this process 1000 times for each value of $p$.
In the following table, we present the mean and standard deviation of the mean
(1000 simulated realizations of $p_1,\ldots,p_N$) of the expected total number of tests
for $D^{'}, S$, HDP, and CRC for population sizes of 20 and 100. In addition, we present the lower bound (Shannon entropy $H(p)$) of the optimal group testing algorithm among the general class
(see discussion and references in \cite{M2017}).
For $N=20$, we performed CRC with four stages since the performance of the algorithm was better than that with three stages in this case.
For $N=100$, we performed CRC with three stages since we were not be able to execute the CRC with four stages even with single realization, while HDP requires less than 1 second (Lenovo X1 Carbon, Intel Core i5-7200U 7th Gen, 2.50HGz, 8GB Ram). The running time for the CRC algorithm with four stages is increasing exponentially with $N$.
For a particular implementation, $p=0.01$, the computational time was 0.85 and 3195 seconds for $N=10$ and $N=60$, respectively. The computational time increased by a factor of about 5 for each increase in $N$
of 10.

Table \ref{tg:1} shows that particularly for large $N$, HDP is substantially more efficient (expected number of tests) than CRC. Furthermore, the HDP algorithm is efficient relative
to the unattainable lower bound $H(p)$ for all but very small $p$. Interestingly, both of the standard procedures $D^{\prime}$ and $S$ outperform CRC when $p\geq 0.05$ and $N=100$,
and procedure $S$ outperforms CRC when $N=20$ for $p\geq 0.05$.
Additionally, for $p\geq 0.20$ the expected total number of tests for CRC is greater than 100, showing that individual testing is more efficient than CRC in this case.

\begin{table}[H]
\caption{
The mean (standard deviation of the mean)
(based on 1000 simulated realizations of $p_1,\ldots,p_N$) of the expected total number of tests
for procedures $D^{'},\,S,\,HDP,$ and $CRC$. The ratio $R$ of the expectation of $HDP$ to $CRC$ and the lower bound $H(p)$ is also presented.
}
\label{tg:1}
\small
\begin{center}
  \begin{tabular}{llSSSSSSS}
    \toprule
    \multirow{1}{*}{$p$} &
      \multicolumn{6}{c}{$N=20$ } \\
      &{$D^{'}$} & {$S$} &{$HDP$} & {$CRC$} & {$R$} &  {$H(p)$}\\
      \midrule
 0.001    &{1.390 (0.003) }   &{1.178 (0.001) }   &{1.108 (0.001) }   &{1.146 (0.001) }  & {0.967}   &{0.215 (0.001) } \\
 0.010    &{3.624 (0.012) }   &{2.701 (0.012) }   &{2.052 (0.008) }   &{2.397 (0.010) }  & {0.856}   &{1.488 (0.009) } \\
 0.050    &{7.523 (0.024) }   &{6.357 (0.024) }   &{5.663 (0.026) }   &{6.871 (0.032) }  & {0.824}   &{5.123 (0.025) } \\
 0.100   & {10.147 (0.031)}   &{9.126 (0.033) }   &{8.656 (0.036) }  & {10.512 (0.042)}   &{0.823}   &{8.088 (0.035) } \\
 0.200   & {13.479 (0.035)}   &{12.784 (0.039)}   &{12.510 (0.041)}   &{14.715 (0.044)}   &{0.850}  &{11.937 (0.041) } \\
 0.300   & {15.580 (0.034)}   &{15.120 (0.037)}   &{14.932 (0.040)}   &{17.728 (0.069)}   &{0.842}  &{14.082 (0.039) } \\
       \bottomrule
  \end{tabular}
  \smallskip

   \begin{tabular}{llSSSSSSS}
    \toprule
    \multirow{1}{*}{$p$} &
      \multicolumn{6}{c}{$N=100$ } \\
      &{$D^{'}$} & {$S$} &{$HDP$} & {$CRC$} & {$R$} &  {$H(p)$}\\
      \midrule
  0.001   &{5.727 (0.009) }   &{3.737 (0.006) }   &{1.864 (0.003) }    &{2.740 (0.005)  }   & 0.680   &{1.078 (0.003) } \\
  0.010   &{17.297 (0.027)}   &{13.080 (0.023)}   &{8.788 (0.021) }    &{14.923 (0.034) }   & 0.589   &{7.437 (0.019) } \\
  0.050   &{37.086 (0.055)}   &{31.793 (0.055)}   &{28.283 (0.059)}    &{39.021 (0.060) }   & 0.725   &{25.647 (0.057)}  \\
  0.100   &{50.733 (0.071)}   &{46.080 (0.075)}   &{43.714 (0.081)}    &{99.885 (0.235) }   & 0.438   &{40.835 (0.080)}  \\
  0.200   &{67.430 (0.080)}   &{64.209 (0.087)}   &{62.935 (0.092)}    &{101.000 (0.000)}   & 0.623   &{59.978 (0.092)}  \\
  0.300   &{77.665 (0.076)}   &{75.433 (0.083)}   &{74.662 (0.087)}    &{101.000 (0.000)}   & 0.739   &{70.383 (0.086)}  \\
     \bottomrule
  \end{tabular}
  \end{center}
  \end{table}

The same numerical comparisons under a non-differential misclassification assumption are presented in Appendix \ref{se:IT},
with corresponding technical details in Appendix \ref{A:M}.

\section{Screening analysis for oral HPV}
\label{se:HPV}
We evaluate different designs for oral HPV screening using the NHANES cohort
from 2011-2012. The publicly available NHANES data contain screening
information on 37 HPV subtypes for men and women ages 18-69 (the full list is available at
\url{https://wwwn.cdc.gov/Nchs/Nhanes/2011-2012/ORHPV_G.htm})
, where
HPV is defined as being positive on any subtype.
Table 2 shows individual characteristics for participant in this cohort, including 8.1\%
overall HPV prevalence.

\begin{table}[]
\label{tg:D}
\caption{Basic summary of the prediction model variables for the 2011-2012 NHANES cohort, which is used for illustration.}
\centering
\begin{tabular}{lc}
\toprule
    N   &  3883    \\
    Gender   &       \\
    \qquad Female   &  1886 (48.6\%)    \\
    \qquad Male   &  1997 (51.4\%)    \\
    Age (mean (se))   &  42.10 (15.25)     \\
    Ethnicity   &       \\
    \qquad African American   &  1101 (28.4\%)    \\
    \qquad Caucasian   &  1333 (34.3\%)    \\
    \qquad Mexican American   &  413 (10.6\%)    \\
    \qquad Other   &  635 (16.4\%)    \\
    \qquad Other Hispanic   &  401 (10.3\%)    \\
    Smoker   &       \\
    \qquad Current ($<$ 10/day)   &  574 (14.8\%)    \\
    \qquad Current ($>$ 10/day, $\leq$ 20/day)   &  238 (6.1\%)    \\
    \qquad Current ($>$ 20/day)   &  74 (1.9\%)    \\
    \qquad Never/Former   &  2997 (77.2\%)    \\
    Lifetime Partners   &       \\
    \qquad 1   &  519 (13.4\%)    \\
    \qquad 11-20   &  555 (14.3\%)    \\
    \qquad 2-5   &  1204 (31.0\%)    \\
    \qquad 6-10   &  793 (20.4\%)    \\
    \qquad $>$ 20   &  545 (14.0\%)    \\
    \qquad None   &  267 (6.9\%)    \\
    HPV Positive   &  315 (8.1\%)    \\
\bottomrule
\end{tabular}
\end{table}

We develop a prediction model using individual-specific characteristics that are known to affect HPV prevalence in order to choose the design configuration ($p_i$'s). A logistic regression model was fit
to an earlier NHANES cohort (2009-2011) with covariates chosen based on
the findings of \cite{GBPWKGC2012} and included gender, age,
ethnicity, smoking status, and the lifetime number of sexual partners.

Figure 1 shows the initial group size distribution as determined by the
HDP algorithm.  The group sizes range from 1 to 89, with a total of 442
groups among  3883 individuals. We use this initial group
configuration for the CRC algorithm, since this method requires that initial
group sizes be specified, and no guidance for large $N$ is provided \citep{BBT2015}.
\begin{figure}[H]
\label{fig:unnamed-chunk-6}
\centering
\includegraphics[width = .5\linewidth]{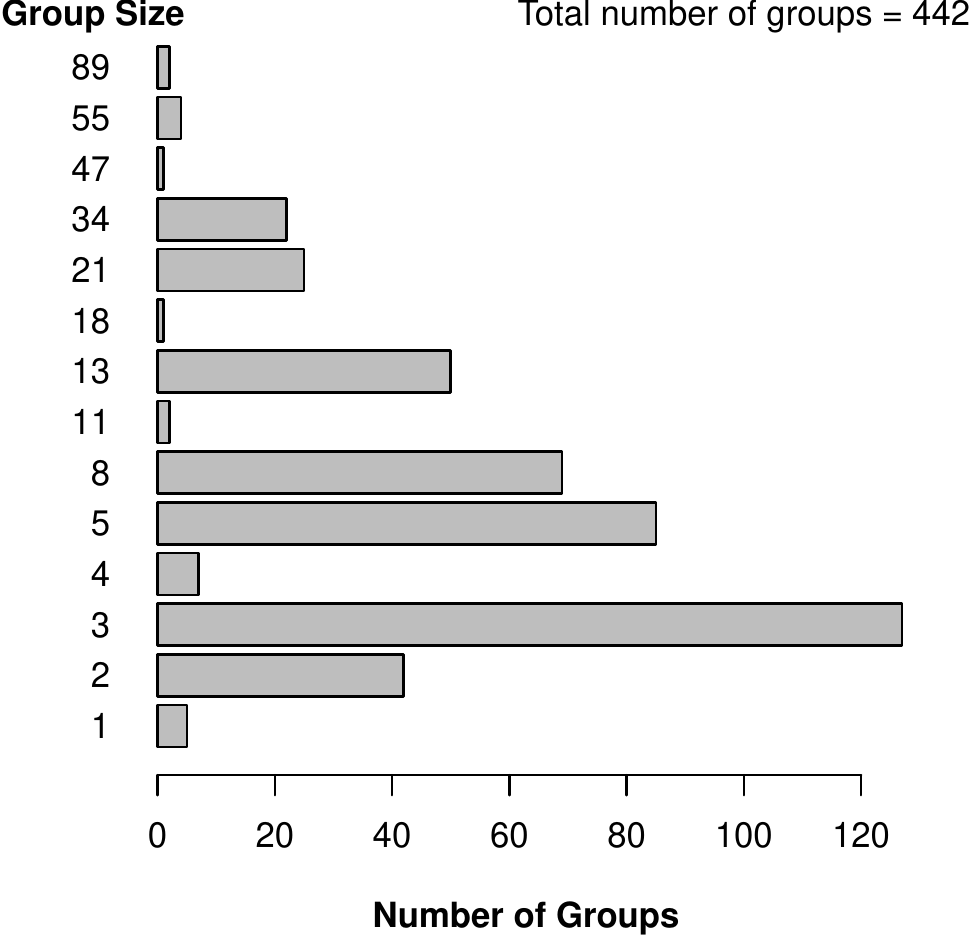}
\caption{Initial group sizes as determined by the HDP algorithm.}
\end{figure}
We compare the total number of tests required to screen the population using the HDP and CRC algorithms in Table 3.
We used a 3-stage CRC algorithm since a 4-stage algorithm is not computationally feasible with this population size and first-stage configuration (Figure 1). The HDP algorithm shows a 16\% (1-1588/1894) efficiency gain relative to the CRC procedure.
In addition, we executed the CRC procedure with three stages (CRC*(3S)) and four stages (CRC*(4S)) by grouping all subjects into groups of size 20 as done by \cite{BBT2015}. This latter first-stage configuration for the CRC procedure
is less efficient than using the HDP initial configuration (i.e., CRC has a smaller number of tests than CRC*(3S) and CRC*(4S) ).

\begin{table}[H]
\label{tab:example}
\caption{Comparison of HDP with CRC.
CRC uses the same initial configuration as HDP with a maximum of three stages.
The initial configuration for three stages (CRC*(3S)) and four stages (CRC*(4S)) was found by grouping all subjects into groups of size 20.
}
\centering
\begin{tabular}{cc}
\toprule
 \multicolumn{1}{c}{} & \multicolumn{1}{c}{Number of Tests} \\
    HDP   &  1588   \\
    CRC   &  1894    \\
    CRC*(3S)   & 2011     \\
    CRC*(4S)   & 1907   \\
\bottomrule
\end{tabular}
\end{table}
An analysis that compares the HDP and CRC designs for oral HPV under non-differential misclassification is presented in Appendix \ref{se:IT}.

\section{Discussion}
\label{se:D}
This article presents an optimal design strategy (HDP) for hierarchical generalized group testing.
We compared the performance of the HDP and CRC approaches, and showed a marked improvement in the efficiency of HDP.
As compared with CRC, the HDP is substantially more efficient in terms of both the expected total number of tests and computational feasibility. In fact, in our experience, the CRC could not be executed
in population sizes larger than 100 with prevalences that are not extremely small ($p_i$'s$\geq 0.01$).

The maximum number of stages in the HDP algorithm can be determined based on a function of the individual prevalences.
Therefore, the maximum number of stages can be determined before beginning the screening protocol since the individual prevalences
will be specified. However, if we want to limit the maximum number of stages (without respect to population size and individual prevalences), deriving the optimal design subject to an arbitrary maximum number of stages is a very difficult optimization problem. If the practitioner needs to restrict the maximum number of stages to a value $k$ that is less than the one corresponding to an optimal design, she/he can use the design specified by the first $k-1$ stages of an optimal design followed by individual testing in the last stage.

There are practical considerations that need to be addressed in designing screening studies using group testing.
Large scale screening programs may be implemented across large geographical areas.
In these situations, there may be a choice between implementing the algorithm across the entire population or within smaller geographic areas. The former design would be more efficient. However, practical considerations (e.g., combining samples from different geographic regions may be logistically difficult) may make implementing the algorithm within regions more advantageous.

We examined the performance of all of the discussed algorithms for the case of imperfect tests under the non-differential misclassification assumption in Appendices \ref{A:M} and \ref{se:IT}. However, we recognize that the optimality criterion of minimizing the expected number of tests is problematic since the classification ability needs to be considered.
To be specific, group testing for screening under misclassification should consider an objective function that factors in the overall sensitivity and specificity in addition to the expected total number of tests \citep{GR1972,H1976,MAR2016}.
Future research should focus on the optimal design of hierarchical generalized group testing with erroneous tests that incorporates misclassification into the design criterion.

\section*{Acknowledgments}
The authors thank the editors, the associate editor and the referees for their thoughtful and constructive comments and suggestions.

\bigskip
\begin{center}
{\large\bf Appendices}
\end{center}

\begin{description}
\item[Appendix A] Development of an optimal HDP algorithm for GGTP
\item[Appendix B] Hierarchical Algorithm to minimize Expected number of Tests in the presence of non-differential misclassification
\item[Appendix C] Design considerations with imperfect tests

\end{description}

\section*{Appendix}
\appendix
\section{Development of an optimal HDP algorithm for GGTP}
\label{A:aa}
\medskip
We show here the development of an optimal HDP for GGTP. As we already discussed in Section \ref{se:OHDP},
we impose an order restriction $p_1\leq p_2\leq \cdots \leq p_N$.
In the homogeneous case, i.e., $p_1=p_2=\ldots=p_N=p$ an optimal hierarchical DP algorithm was obtained by \cite{SG1959}
and recently rediscovered and computationally improved by \cite{Z2017} (see also \cite{M2019} for the discussion).

\subsection {Evaluation of $H(n:N)$}
\label{A1:H}
Recall that we are dealing with the binomial set $B_{n:N}$.
We begin with the case $n=N$. In this case $H(N:N)=1$.
For subsequent evaluation, when $n\leq N-1$,
we have to find the size of $k$ of the subset $B_{n:n+k-1}$ from $B_{n:N}$ to test. If the test outcome of $B_{n:n+k-1}$ is negative, then we test the remaining units $n+k,\ldots,N$ that form a binomial set $B_{n+k:N}$.
Otherwise, if the test outcome of $B_{n:n+k-1}$ is positive, then its units form the defective set of size $k$, which we abbreviate as $D_{n:n+k-1}$, and remaining units from $B_{n:N}$ form the binomial set $B_{n+k:N}$.
We summarize these situations in the following binary testing tree.
Recall that, the left branch of the tree represents a negative test result, and the right branch
represents a positive test result.

\begin{figure}[H]
\Tree [.{test $B_{n:n+k-1}\left(\subseteq  B_{n:N}\right)$} [{$B_{n+k:N}$ with prob. $q_n\cdots q_{n+k-1}$} ] [{$D_{n:n+k-1}\cup B_{n+k:N}$ with prob. $1-q_n\cdots q_{n+k-1}$} ] ]
\label{fig:A}
\end{figure}

Denote by $T_{k}(n:N)$ the expected total number of tests. Then,
\begin{align}
\label{eq:TH}
&
T_{k}(n:N)=1+\left\{  q_n\cdots q_{n+k-1} H(n+k:N)\nonumber
+\left(1-q_n\cdots q_{n+k-1}\right)
\left[h(n:n+(k-1))+H(n+k:N)\right]\right\}\\
&
=
1+\left\{H(n+k:N)+\left(1-\Pi\left(n:n+k-1\right)\right)h(n:n+(k-1))\right\},
\end{align}
where
$\displaystyle \Pi\left(n:n+k-1\right)=q_n\cdots q_{n+k-1}. $\\
Since the optimal value $H(n:N)$ is obtained by choosing the best $k$ among $k=1,\ldots,N-(n-1)$, we have
\begin{equation}
\label{eq:H}
H(n:N)=\min_{1\leq k \leq N-(n-1)}T_{k}(n:N), \,\,\,n=N-1,\ldots,1.
\end{equation}
Then $H(n:N)$ is calculated in a recursive manner for $n=N-1,\ldots,1$.
This calculation \eqref{eq:TH} required the conditional expectation $h()$, which is developed as follow.

\subsection{Evaluation of $h(n:n+(l-1))$}
\label{A2:h}
Recall that we are dealing with the defective set $D_{n:n+(l-1)}$.
If $l=1$, then $h(n:n+(l-1))=h(n:n)=0$. If $l\geq 2$, we have to find
a proper subset $\{n:n+x-1\}$ of size $x$ from $D_{n:n+(l-1)}$ to test.
If the binary test outcome of $\{n:n+x-1\}$ is negative, then we conclude that the remaining units $n+x,\ldots,n+(l-1)$ form a defective set $D_{n+x:n+(l-1)}$, which does not need to be tested as a whole set.
If the test outcome of $\{n:n+x-1\}$ is positive, then the
conditional posterior distribution
of units $n+x,\ldots,n+(l-1)$ is the same as it was before any testing and they form a binomial set $B_{n+x:n+(l-1)}$
(similar arguments as in \cite{SG1959}).
Therefore, we divide the defective set $D_{n:n+(l-1)}$ into two subsets $\left\{n,n+1,\ldots,n+(x-1)\right\}$
and $\left\{n+x,\ldots,n+(l-1)\right\}$ and test them separately from left to right.
We have three possible states of these subsets, i.e.,  $-+, +-, ++$,
where, for example, $++$ represent the situation where both subsets are positive.
Denote by $T_{ab}$ the expected total number of tests
corresponding to the situation $ab,\,ab\in\{-+, +-, ++\}$.
The following diagram represents all these possible outcomes with corresponding conditional (on the event that there is at least one defective element) probabilities.

\begin{figure}[h!]
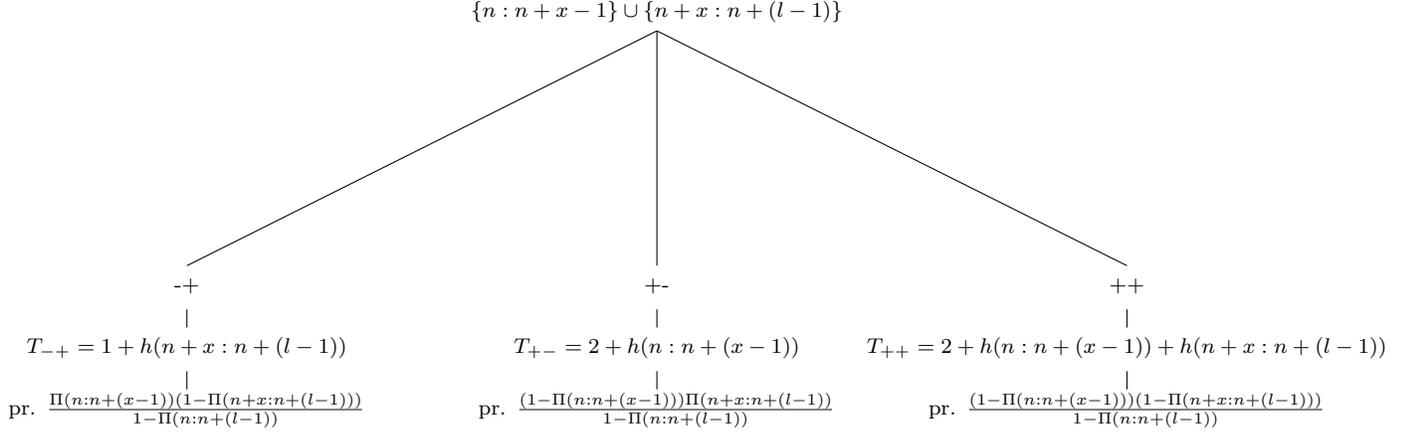

\label{d:1}
\scriptsize
\Tree [
.{$\{n:n+x-1\}\cup \{n+x:n+(l-1)\}$}
[.{-+} [.{$T_{-+}=1+h(n+x:n+(l-1))$}
{pr. $\frac{\Pi\left(n:n+(x-1)\right)\left(1-\Pi\left(n+x:n+(l-1)\right)\right)}{1-\Pi\left(n:n+(l-1)\right)}$}
 ]]
[.{+-} [.{$T_{+-}=2+h(n:n+(x-1))$}
{pr. $\frac{\left(1-\Pi\left(n:n+(x-1)\right)\right)\Pi\left(n+x:n+(l-1)\right)}{1-\Pi\left(n:n+(l-1)\right)}$}
]]
[.{++} [.{$T_{++}=2+h(n:n+(x-1))+h(n+x:n+(l-1))$}
{pr. $\frac{\left(1-\Pi\left(n:n+(x-1)\right)\right)\left(1-\Pi\left(n+x:n+(l-1)\right)\right)}{1-\Pi\left(n:n+(l-1)\right)}$}
]]
 ]
\caption{Possible outcomes with corresponding conditional (on the event that there is at least one defective element) probabilities.}
\label{fig:B}
\end{figure}

Denote by $T\left(n,l,x\right)$ the expected total number of tests in this case. Then,
\begin{align*}
&
T\left(n,l,x\right)\\
&
=\left[1+h(n+x:n+(l-1))\right]
\frac{\Pi\left(n:n+(x-1)\right)\left(1-\Pi\left(n+x:n+(l-1)\right)\right)}{1-\Pi\left(n:n+(l-1)\right)}\\
&
+
\left[2+h(n:n+(x-1))\right]
\frac{\left(1-\Pi\left(n:n+(x-1)\right)\right)\Pi\left(n+x:n+(l-1)\right)}{1-\Pi\left(n:n+(l-1)\right)}\\
&
+
\left[2+h(n:n+(x-1))+h(n+x:n+(l-1))\right]
\frac{\left(1-\Pi\left(n:n+(x-1)\right)\right)\left(1-\Pi\left(n+x:n+(l-1)\right)\right)}{1-\Pi\left(n:n+(l-1)\right)}\\
&
=
2-\frac{\Pi\left(n:n+(x-1)\right)\left(1-\Pi\left(n+x:n+(l-1)\right)\right)}{1-\Pi\left(n:n+(l-1)\right)}
+
\frac{1-\Pi\left(n:n+(x-1)\right)}{1-\Pi\left(n:n+(l-1)\right)}h(n:n+(x-1))\\
&
+
\frac{1-\Pi\left(n+x:n+(l-1)\right)}{1-\Pi\left(n:n+(l-1)\right)}h(n+x:n+(l-1)).
\end{align*}
Since an optimal value $h(n:n+(l-1))$ is obtained by choosing the best $x$, among $x=1,\dots,l-1$ we have,
\begin{equation}
\label{eq:h}
h(n:n+(l-1))=\min_{1\leq x \leq l-1}T\left(n,l,x\right).
\end{equation}

Combining \eqref{eq:H} and \eqref{eq:h}, we obtain an optimal ordered HDP algorithm:
\begin{center}
\begin{align}
\label{eq:DP}
&
\displaystyle{ H(N+1:N)=0,\,H(N:N)=1,\, h(n:n)=0,\, n=1,\ldots,N }\\\nonumber
&
\displaystyle
{
H(n:N)=1+\min_{1\leq k \leq N-(n-1)}\left\{H(n+k:N)+\left(1-\Pi\left(n:n+(k-1)\right)\right)h(n:n+(k-1))\right\},
}
\\
\nonumber
&
\displaystyle
{
h(n:n+(l-1))=2+\min_{1\leq x \leq l-1}\Big\{-\frac{\Pi\left(n:n+(x-1)\right)\left(1-\Pi\left(n+x:n+(l-1)\right)\right)}{1-\Pi\left(n:n+(l-1)\right)}
}
\\\nonumber
&
+
\displaystyle
{
\frac{1-\Pi\left(n:n+(x-1)\right)}{1-\Pi\left(n:n+(l-1)\right)}h(n:n+(x-1))+
\frac{1-\Pi\left(n+x:n+(l-1)\right)}{1-\Pi\left(n:n+(l-1)\right)}h(n+x:n+(l-1))
\Big\},
}\\\nonumber
&
n=N-1,N-2,\ldots,1;\\\nonumber
&
l=2,\ldots, N-n+1,
\end{align}
\end{center}
where $\displaystyle \Pi\left(a:b\right)=q_{a}q_{a+1}\cdots q_{b}$.

\section{Hierarchical algorithm to minimize expected number of tests in the presence of non-differential misclassification}
\label{A:M}
In this section we extend the algorithm presented in Appendix \ref{A:aa} to the case where tests are subject to the non-differential misclassification. It can be done in a straightforward manner by re-calculating $H(n:N)$ and $h(n:n+(l-1))$ from Appendix \ref{A:aa}.
\subsection{Evaluation of $H_{M}(n:N)$}
\label{sse:1}
\begin{equation}
\label{eq:TkM}
T_{k,M}(n:N)
=
1+\left\{H_{M}(n+k:N)+\left(1-\Pi_{M}\left(n:n+k-1\right)\right)h_M(n:n+(k-1))\right\},
\end{equation}
where $\Pi_{M}\left(n:n+k-1\right)=S_p\Pi\left(n:n+k-1\right)+(1-S_e)(1-\Pi\left(n:n+k-1\right)),\\
\Pi\left(n:n+k-1\right)=q_{n}\cdots q_{n+k-1}$.
\begin{equation}
\label{eq:HM}
H_{M}(n:N)=\min_{1\leq k \leq N-(n-1)}T_{k, M}(n:N), \,\,\,n=N-1,\ldots,1.
\end{equation}

\subsection{Evaluation of $h_{M}(n:n+(l-1))$}
\label{sse:2}
Given that the test outcome of the items $n:n+l-1$ is positive, we have to calculate the below
probabilities, which correspond to all possible situations.

\begin{align*}
&
P_{-}=P(\text{test outcome of $n:n+x-1$ is negative}\,|\,\text{test outcome of $n:n+l-1$ is positive})\\
&
= \frac{S_e(1-S_e)+\Pi_{a}\left(S_p S_e-S_e(1-S_e)\right)
+\Pi_{ab}\left(S_p-S_p^2-S_p S_e\right)
}{1-\Pi_{M}\left(n:n+l-1\right)},
\end{align*}

\begin{align*}
&
P_{+-}=P(\text{test outcome of $n:n+x-1$ is + $\cap$ $n+x:n+l-1$ is -}\,|\,\text{test outcome of $n:n+l-1$ is +})
\\
&
=
\frac{S_e^2(1-S_e)+\Pi_a\left(S_e(1-S_e)(1-S_p-S_e)\right)
+\Pi_{b}\left(S_e^2(S_p+S_e-1)\right)
}{1-\Pi_{M}\left(n:n+l-1\right)}
\\
&
+
\frac{\Pi_{ab}\left(S_e(1-S_e)(S_e+S_p-1)+S_p(1-S_p)^2-S_e^2S_p\right)
}{1-\Pi_{M}\left(n:n+l-1\right)},
\end{align*}

\begin{align*}
&
P_{++}=
P(\text{test outcome of $n:n+x-1$ is + $\cap$ $n+x:n+l-1$ is +}\,|\,\text{test outcome of $n:n+l-1$ is +})
\\
&
=
\frac{S_e^3+\Pi_a\left(S_e^2(1-S_p-S_e)\right)
+\Pi_{b}\left(S_e^2(1-S_p-S_e)\right)
}{1-\Pi_{M}\left(n:n+l-1\right)}
\\
&
+
\frac{\Pi_{ab}\left((1-S_p)^3-2S_e^2(1-S_p)+S_e^3\right)
}{1-\Pi_{M}\left(n:n+l-1\right)},
\end{align*}

We summarize the above probabilities with the corresponding total number of the tests in Figure \ref{fig:BM} below.

\begin{figure}[h!]
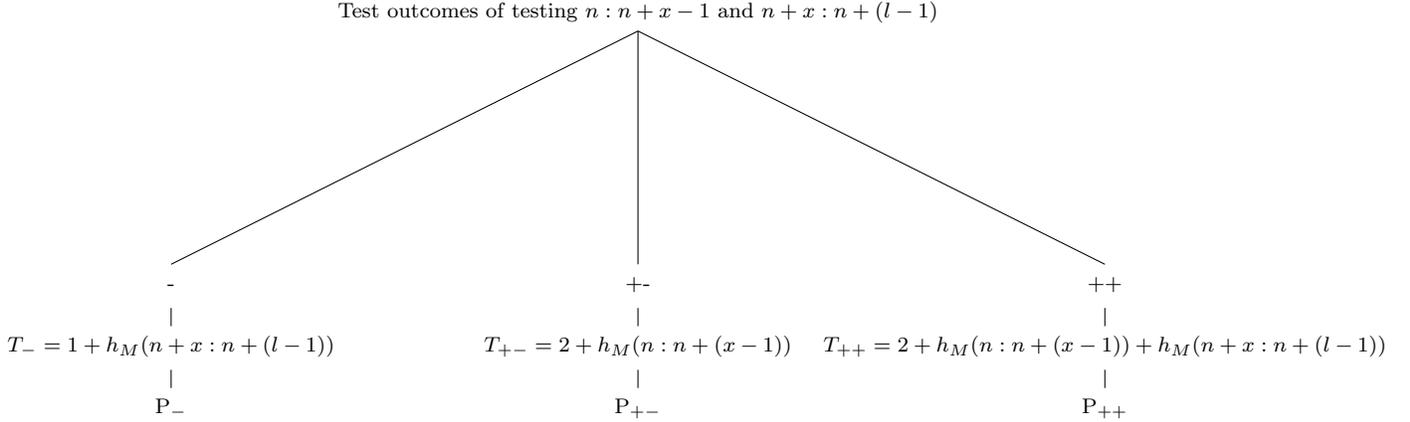

\scriptsize
\Tree [
.{Test outcomes of testing ${n:n+x-1}$ and ${n+x:n+(l-1)}$}
[.{-} [.{$T_{-}=1+h_{M}(n+x:n+(l-1))$}
{P_{-}}
 ]]
[.{+-} [.{$T_{+-}=2+h_{M}(n:n+(x-1))$}
{P_{+-}}
]]
[.{++} [.{$T_{++}=2+h_{M}(n:n+(x-1))+h_{M}(n+x:n+(l-1))$}
{P_{++}}
]]
 ]
\caption{Possible outcomes with corresponding conditional (on the event that test outcome of the items $\left\{n,\ldots,n+l-1\right\}$ is positive) probabilities. }
\label{fig:BM}
\end{figure}

where $\Pi_a=\Pi\left(n:n+x-1\right),\, \Pi_b=\Pi\left(n+x:n+l-1\right),\, \Pi_{ab}=\Pi\left(n:n+l-1\right)$.
We have
$P_{-}+P_{+-}+P_{++}=1$
and
\begin{equation}
\label{eq:TM}
T_{M}\left(n,l,x\right)=T_{-}\times P_{-}+T_{+-}\times P_{+-}+T_{++} \times P_{++}.
\end{equation}

\begin{remark}[$h_{M}$]
Probabilities $P_{-}, P_{+-}, P_{++}$ were calculated by conditioning on the true status
of subsets $n:n+x-1$ and $n+x:n+l-1$, i.e., on 4 possibilities.
Given that the test outcome of the set $n:n+l-1$ of size $l$ is positive we proceed in the following way:
\begin{itemize}
\item
if the test outcome of its proper subset $n:n+x-1$ of size $x$ is negative, then we do not test
the remainder subset $n+x:n+l-1$ and conclude that it is positive. Therefore, the left branch
of the tree has $T_{-}=1+h_{M}(n+x:n+(l-1))$ number of tests.
\item
otherwise, if the test outcome of its proper subset $n:n+x-1$ of size $x$ is positive,
then we test the remainder subset $n+x:n+l-1$ of size $l-x$ and
\begin{itemize}
\item
if its test outcome is negative, then we have $T_{+-}=2+h_{M}(n:n+(x-1))$ tests to perform.
\item
otherwise, if its test outcome is positive, then we have $T_{++}=2+h_{M}(n:n+(x-1))+h_{M}(n+x:n+(l-1))$ tests to perform.
\end{itemize}
\end{itemize}
\end{remark}

\begin{equation}
\label{eq:hM}
h_{M}(n:n+(l-1))=\min_{1\leq x \leq l-1}T_M\left(n,l,x\right).
\end{equation}

\subsection{ An optimal HDP algorithm with respect to the expected total number of tests}
In the Algorithm (Section \ref{se:OHDP}), change $T(n,k,x)$ into $T_{M}(n,k,x)$ from \eqref{eq:TM}.
This will allow calculation of $h_{M}(n:n+(l-1))$ in \eqref{eq:hM}, which will replace $h(n:n+(l-1))$ in the Algorithm (Section \ref{se:OHDP}). Then in the Algorithm (Section \ref{se:OHDP}), change $T_{k}(n:N)$ into $T_{k,M}(n:N)$ (equation \eqref{eq:TkM}).
This will allow calculation of $H_{M}(n:N)$ in \eqref{eq:HM}, which will replace $H(n:N)$ in the Algorithm (Section \ref{se:OHDP}).

\section{Design considerations with imperfect tests}
\label{se:IT}
\cite{BBT2015} incorporated non-differential misclassification (not dependent on prevalence or group sizes)
for the CRC design and used the expected total number of tests as a design criterion.
In a similar setting, we found the HDP design that minimizes the expected total number of tests (Appendix  \ref{A:M}).

We compare HDP and CRC in the setting where tests are subject to misclassification.
We follow  \cite{BBT2015} and assume that test sensitivity (Se) and specificity (Sp) are known.
The comparison was done by executing the design (Appendix \ref{A:M}), formulating the groups, and then misclassifying either the group or individual tests with sensitivity Se and specificity Sp.
Table 4 shows the comparison of CRC and HDP with respect to the expected total number of tests and overall sensitivity (SE) and specificity (SP) for the two population sizes considered in Table \ref{tg:1} (N=20 and 100). The HDP algorithm
was more efficient than CRC among all choices of $p$ and for both population sizes.
The increased efficiency was particulary notable for N=100 for all but $p=0.05$, where HDP has an expected number of tests that is less than half of those with CRC. That said, the sensitivity for HDP is substantially lower than that for CRC in the
case of the larger population size.
The reason is that in order to minimize the expected total number of tests under misclassification, the HDP algorithm tends to form groups that are as large as possible, thus creating very low overall sensitivity. This illustrates the danger in using the expected number of tests as an optimality criterion when tests are imperfect \citep{GR1972, MAR2016}.

To avoid the large group sizes for the HDP algorithm that accounts for misclassification, we used the design computed with DP assuming no misclassification. We subsequently evaluated this design under misclassification (HDP$^{\star}$).
Generally, this approach showed a smaller expected total number of tests with similar overall sensitivity and specificity
as compared with CRC.

\begin{table}[]
\centering
\scriptsize
\caption{
The mean (standard deviation of the mean)
(based on 1000 simulated realizations of $p_1,\ldots,p_N$) of the expected total number of tests $E(T)$, overall sensitivity (SE), and overall specificity (SP), for the CRC and HDP algorithms. Test sensitivity (Se) and specificity (Sp) are set at
0.95 each.
}
\begin{tabular}{lcccccccc}
\toprule
 \multicolumn{1}{l}{p} & \multicolumn{3}{c}{N = 20}& \multicolumn{3}{c}{N = 100} \\
 \multicolumn{1}{l}{\qquad Method} & \multicolumn{1}{c}{E(T)} & \multicolumn{1}{c}{SE} & \multicolumn{1}{c}{SP}& \multicolumn{1}{c}{E(T)} & \multicolumn{1}{c}{SE} & \multicolumn{1}{c}{SP} \\
    0.001 &  &  &\\
    \qquad CRC   &  1.2570 (0.0010)    &  0.8169 (0.0001)    &  0.9999 (0.0000)&  3.6008 (0.0062)    &  0.8574 (0.0000)    &  0.9992 (0.0000)     \\
    \qquad HDP   &  1.1124 (0.0005)    &  0.7616 (0.0012)    &  0.9971 (0.0000)   &  1.5524 (0.0019)    &  0.4797 (0.0018)    &  0.9986 (0.0000)      \\

    \qquad HDP*   &  1.2913 (0.0013)    &  0.8738 (0.0008)    &  0.9973 (0.0000)   &  2.1916 (0.0030)    &  0.8298 (0.0004)    &  0.9992 (0.0000)       \\
    0.010   &      &      &       \\
    \qquad CRC   &  2.4794 (0.0092)    &  0.8227 (0.0003)    &  0.9989 (0.0000)   &  16.8756 (0.0358)    &  0.8574 (0.0000)    &  0.9956 (0.0000)     \\
    \qquad HDP   &  1.9053 (0.0069)    &  0.7969 (0.0007)    &  0.9947 (0.0000)  &  7.6063 (0.0207)    &  0.7372 (0.0003)    &  0.9958 (0.0000)     \\
    \qquad HDP*   &  2.2004 (0.0073)    &  0.8709 (0.0003)    &  0.9956 (0.0000)  &  9.0722 (0.0248)    &  0.8258 (0.0004)    &  0.9966 (0.0000)        \\
    0.050   &      &      &       \\
    \qquad CRC   &  6.7019 (0.0306)    &  0.8370 (0.0003)    &  0.9952 (0.0000)  &  39.7957 (0.0549)    &  0.8579 (0.0000)    &  0.9901 (0.0000)     \\
    \qquad HDP   &  5.3465 (0.0270)    &  0.8406 (0.0003)    &  0.9890 (0.0000)    &  26.2597 (0.0487)    &  0.6779 (0.0012)    &  0.9900 (0.0000)       \\
    \qquad HDP*   &  5.6745 (0.0273)    &  0.8726 (0.0004)    &  0.9893 (0.0001) &  28.5490 (0.0601)    &  0.8736 (0.0001)    &  0.9893 (0.0000)         \\
    0.100   &      &      &       \\
    \qquad CRC   &  10.2127 (0.0383)    &  0.8499 (0.0003)    &  0.9908 (0.0001)   &  86.9833 (0.4329)    &  0.8594 (0.0000)    &  0.9590 (0.0003)        \\
    \qquad HDP   &  8.4694 (0.0356)    &  0.8757 (0.0004)    &  0.9836 (0.0001)  &  33.6802 (0.0222)    &  0.4458 (0.0009)    &  0.9869 (0.0000)     \\
    \qquad HDP*   &  8.7259 (0.0348)    &  0.8949 (0.0003)    &  0.9835 (0.0001) &  43.8472 (0.0766)    &  0.8953 (0.0001)    &  0.9835 (0.0000)         \\
    0.200   &      &      &       \\
    \qquad CRC   &  14.1723 (0.0408)    &  0.8684 (0.0003)    &  0.9833 (0.0001)   &  92.2475 (0.0000)    &  0.8590 (0.0000)    &  0.9549 (0.0000)       \\
    \qquad HDP   &  12.3358 (0.0415)    &  0.9113 (0.0003)    &  0.9762 (0.0001)    &  36.3820 (0.0082)    &  0.3685 (0.0005)    &  0.9873 (0.0000)      \\
    \qquad HDP*   &  12.5043 (0.0406)    &  0.9194 (0.0002)    &  0.9758 (0.0001)    &  62.8737 (0.0923)    &  0.9198 (0.0001)    &  0.9758 (0.0000)       \\
    0.300   &      &      &       \\
    \qquad CRC   &  16.5500 (0.0371)    &  0.8585 (0.0003)    &  0.9772 (0.0001)  &  92.2475 (0.0000)    &  0.8587 (0.0000)    &  0.9549 (0.0000)      \\
    \qquad HDP   &  14.7381 (0.0413)    &  0.9276 (0.0002)    &  0.9710 (0.0001)   &  37.0501 (0.0046)    &  0.3401 (0.0004)    &  0.9892 (0.0000)        \\
    \qquad HDP*   &  14.8583 (0.0403)    &  0.9324 (0.0002)    &  0.9706 (0.0001) &  74.4149 (0.0874)    &  0.9329 (0.0001)    &  0.9708 (0.0000)    \\
\bottomrule
\end{tabular}
\end{table}

In Section \ref{se:HPV}, we evaluated designs for oral HPV under the assumption that tests are error-free.
We now consider screening designs assuming non-differential misclassification with both sensitivity and specificity equal to 0.95 \citep{H2018}.
This was done by executing the design, formulating the groups, and then misclassifying either the group or individual tests with sensitivity and specificity of 0.95. Table 5 shows the total number of tests as well as overall sensitivity and specificity for the CRC and HDP algorithms.
The HDP algorithm is much more efficient than CRC with respect to the expected total number of tests
but has very low overall sensitivity.
As explained above, this is due to the tendency to form large groups in order to minimize the total number of tests. This very small number of tests and low sensitivity demonstrate the danger of simply using the total number of tests as a design criterion under misclassification. As recommended earlier, we used the HDP algorithm without misclassification for design purposes (HDP*).
The HDP* algorithm results in a larger total number of tests than HDP, but with reasonable overall sensitivity. In fact, this design resulted in a smaller number of tests than using CRC (14\% reduction), but
with similar overall sensitivity and specificity.
In addition, we follow \cite{BBT2015} (Section 5, data example) and execute the CRC algorithm with a maximum of three stages  (CRC* (3S)) and four stages ( CRC* (4S)) by grouping all subjects into groups of size 20 in the first stage.
In this case, CRC* (3S) has a similar overall sensitivity and specificity but is inefficient relative to HDP*. The CRC* (4S) procedure is more efficient than CRC* (3S) and has lower overall sensitivity than HDP*.

\begin{table}[H]
\caption{Results from screening algorithms with both sensitivity and specificity equal 0.95}
\centering
\begin{tabular}{llll}
\toprule
 \multicolumn{1}{c}{} & \multicolumn{1}{c}{Number of Tests} & \multicolumn{1}{c}{SE} & \multicolumn{1}{c}{SP} \\
    HDP   &  10   &  0.010   &  1.000    \\
    HDP*   &  1606   &  0.895   &  0.985    \\
    CRC   &  1877   &  0.870   &  0.994    \\
    CRC* (3S)&	2031&	0.889&	0.987\\
    CRC* (4S)&	1778&	0.797&	0.992\\
\bottomrule
\end{tabular}
\end{table}

Our analysis using both the HPD and CPC shows the advantages of the HPD in this setting. However, it also shows the problem with using the expected number of tests as an optimality criterion under testing error.

{}


\begin{thebibliography}{}

\bibitem[\protect\citeauthoryear{Bellman}{1957}]{B1957}
Bellman, R. (1957).
\newblock Dynamic Programming. {\emph Princeton University Press}.

\bibitem[\protect\citeauthoryear{Bilder \it{et~al.}}{2010}]{BT2010}
Bilder, C.~R., Tebbs, J.~M., Chen, P. (2010).
\newblock Informative retesting.
\newblock {\emph J. Am. Stat. Assoc.} {\textbf 105,} 942--955.

\bibitem[\protect\citeauthoryear{Black \it{et~al.}}{2015}]{BBT2015}
Black, M.~S., Bilder, C.~R., Tebbs, J.~M. (2015).
\newblock Optimal retesting configurations for hierarchical
group testing.
\newblock {\emph Appl. Statist.} {\textbf 64,} 693--710.

\bibitem[\protect\citeauthoryear{Dorfman}{1943}]{Dorfman1943}
Dorfman, R. (1943).
\newblock The detection of defective members of large populations.
\newblock {\emph The Annals of Mathematical Statistics } {\textbf 14,} 436--440.



\bibitem[\protect\citeauthoryear{Gillison \it{et~al.}}{2012}]{GBPWKGC2012}
Gillison, M. L., Broutian, T., Pickard, R.K., Tong, Z.Y., Xiao, W., Kahle, L., Graubard, B. I., Chaturvedi, A. K. (2012).
Prevalence of Oral HPV Infection in the United States, 2009-2010.
{\it JAMA } {\textbf 307,} 693--703.



\bibitem[\protect\citeauthoryear{Graff and Roeloffs}{1972}]{GR1972}
Graff, L. E., and Roeloffs, R. (1972).
Group testing in the presence of test error: an extension of the Dorfman procedure.
{\it Technometrics } {\textbf 14,} 113--122.




\bibitem[\protect\citeauthoryear{Hwang}{1975}]{H1975}
Hwang, F.~K. (1975).
\newblock A generalized binomial group testing problem.
\newblock {\emph J. Amer. Statist. Assoc.} {\textbf 70,} 923--926.

\bibitem[\protect\citeauthoryear{Hwang}{1976}]{H1976}
Hwang, F.~K. (1976).
\newblock Group testing with a dilution effect.
\newblock {\emph Biometrika} {\textbf 63,} 671--673.


\bibitem[\protect\citeauthoryear{Hwang \it{et~al.}}{1981}]{HPE1981}
Hwang, F.~K., Pfeifer, C.~J., and Enis, P. (1981).
\newblock An Optimal Hierarchical Procedure for a Modified Binomial Group-Testing Problem.
\newblock {\emph J. Amer. Statist. Assoc. } {\textbf 76,} 947--949.

\bibitem[\protect\citeauthoryear{Hyun \it{et~al.}}{2018}]{H2018}
Hyun, N., Gastwirth, J. L., Graubard, B. I.  (2018).
\newblock Grouping methods for
estimating prevalences of rare traits for complex survey data that preserve
confidentiality of respondents
\newblock {\emph Statistics in Medicine } {\textbf 37,} 2174--2186.


\bibitem[\protect\citeauthoryear{Lindley}{1961}]{L1961}
Lindley, D.~V. (1961).
\newblock Dynamic Programming and Decision Theory.
\newblock {\emph Appl. Statist.} {\textbf 10,} 39--51.


\bibitem[\protect\citeauthoryear{Litvak \it{et~al.}}{1994}]{LTP1994}
Litvak, E., Tu, X.~M., and Pagano, M. (1994).
\newblock Screening for the Presence of a Disease by Pooling Sera Samples.
\newblock {\emph J. Amer. Statist. Assoc. } {\textbf 89,} 424--434.

\bibitem[\protect\citeauthoryear{Malinovsky \it{et~al.}}{2016}]{MAR2016} Malinovsky, Y., Albert, P. S., and Roy, A. (2016).
Reader Reaction: A Note on the Evaluation of Group Testing
Algorithms in the Presence of Misclassification.
{\it Biometrics} {\textbf 72,} 299--304.

\bibitem[\protect\citeauthoryear{Malinovsky}{2019a}]{M2017}
Malinovsky, Y. (2019a).
\newblock Sterrett procedure for the generalized group testing problem.
\newblock {\emph  Methodology and Computing in Applied Probability}. {\textbf 21,} 829--840.

\bibitem[\protect\citeauthoryear{Malinovsky}{2019b}]{M2019}
Malinovsky, Y. (2019b).
\newblock End Notes.
\newblock {\emph Math. Mag.} {\textbf 92,} 398.

\bibitem[\protect\citeauthoryear{Nanda \it{et~al.}}{2000}]{N2000}
Nanda, K., McCrory, D.~C., Myers, E.~R., Bastian, L.~A., Hasselblad, V., Hickey, J.~D., Matchar, D.~B. (2000).
Accuracy of the Papanicolaou test in screening for and follow-up of cervical cytologic abnormalities: a systematic review.
{\it Ann.Intern.Med.} {\textbf 132,} 810--819.



\bibitem[\protect\citeauthoryear{Schiffman and Wentzensen}{2013}]{SW2013}
Schiffman, M., Wentzensen, N. (2013).
\newblock Human papillomavirus infection and the multistage carcinogenesis of cervical cancer.
\newblock {\it Cancer Epidemiol Biomarkers Prev.} {\textbf 22,} 553--560.

\bibitem[\protect\citeauthoryear{Sobel and Groll}{1959}]{SG1959}
Sobel, M., Groll, P. A. (1959).
\newblock Group testing to eliminate efficiently all defectives in a binomial sample.
\newblock {\it Bell System Tech. J.} {\textbf 38,} 1179--1252.


\bibitem[\protect\citeauthoryear{Sobel}{1960}]{S1960}
Sobel, M. (1960).
\newblock Group testing to classify efficiently all defectives in a binomial sample.
\newblock {\it Information and Decision Processes (R. E. Machol, ed.; McGraw-Hill, New York),} pp. 127-161.

\bibitem[\protect\citeauthoryear{Sterrett}{1957}]{S1957}
Sterrett, A. (1957).
\newblock On the detection of defective members of large populations.
\newblock {\emph The Annals of Mathematical Statistics } {\textbf 28,} 1033--1036.

\bibitem[\protect\citeauthoryear{Zimmerman}{2017}]{Z2017}
Zimmerman, S. (2017).
\newblock Detecting deficiencies: an optimal group testing algorithm.
\newblock {\emph Math. Mag.} {\textbf 90,} 167--178.



\end{thebibliography}
\end{document}